\documentclass[aps,prx,letterpaper,amsmath,amssymb,superscriptaddress,nofootinbib,twocolumn]{revtex4-2}

\usepackage{graphicx}
\usepackage[usenames]{color}
\usepackage[colorlinks=true,linkcolor=blue,citecolor=blue,anchorcolor=green,urlcolor=blue]{hyperref}

\usepackage[percent]{overpic}
\usepackage{tabularx}
\usepackage{multirow}

\usepackage{CJK}

\newcommand{\pll}{\kern 0.56em/\kern -0.8em /\kern 0.56em}

\def\be{\begin{equation}}
\def\ee{\end{equation}}
\def\bea{\begin{eqnarray}}
\def\eea{\end{eqnarray}}

\def\CNO{CoNb$_2$O$_6$}
\def\D81{$\mathcal{D}_8^{(1)}$}
\def\E8{$E_8$}
\begin{document}

\title{Emergent \D81
	spectrum and topological soliton excitation in \CNO}
\author{Ning Xi}
\thanks{These authors contributed equally to the work.}
\affiliation{Department of Physics and Beijing Key Laboratory of Opto-electronic Functional Materials and Micro-nano Devices, Renmin University of China, Beijing 100872, China }

\author{Xiao Wang }
\thanks{These authors contributed equally to the work.}
\affiliation{Tsung-Dao Lee Institute,
Shanghai Jiao Tong University, Shanghai, 201210, China}

\author{Yunjing Gao}
\affiliation{Tsung-Dao Lee Institute,
Shanghai Jiao Tong University, Shanghai, 201210, China}

\author{Yunfeng Jiang}
\affiliation{School of Phyiscs and Shing-Tung Yau Center, Southeast University, Nanjing 210096, China}

\author{Rong Yu}
\altaffiliation{rong.yu@ruc.edu.cn}
\affiliation{Department of Physics and Beijing Key Laboratory of Opto-electronic
Functional Materials and Micro-nano Devices, Renmin University of
China, Beijing 100872, China }
\affiliation{Tsung-Dao Lee Institute,
Shanghai Jiao Tong University, Shanghai, 201210, China}
\affiliation{Key Laboratory of Quantum State Construction and Manipulation (Ministry of Education), Renmin University of China, Beijing, 100872, China}

\author{Jianda Wu}
\altaffiliation{wujd@sjtu.edu.cn}
\affiliation{Tsung-Dao Lee Institute,
Shanghai Jiao Tong University, Shanghai, 201210, China}
\affiliation{School of Physics \& Astronomy, Shanghai Jiao Tong University, Shanghai, 200240, China}
\affiliation{Shanghai Branch, Hefei National Laboratory, Shanghai 201315, China}

\begin{abstract}
Quantum integrability emerging near a quantum critical point (QCP) is
manifested by exotic excitation spectrum that is organised by the
associated algebraic structure. A well known example is the emergent \E8 integrability near the QCP of a transverse field Ising chain (TFIC), which was long predicted theoretically and initially proposed to be
realised in the quasi-one-dimensional (q1D) quantum magnet CoNb$_2$O$_6$.
However, later measurements on the spin excitation spectrum of this material
revealed
a series of satellite peaks
that cannot be 
described by the \E8 Lie algebra.
Motivated by these experimental progresses, we hereby revisit the spin
excitations of CoNb$_2$O$_6$ by combining numerical calculation and analytical analysis.
We show that, as effects of strong interchain fluctuations, the spectrum of the system near the 1D QCP is
characterised by the \D81
Lie algebra with
robust topological soliton excitation.
We further show that the \D81 spectrum can be realised
in a broad class of interacting quantum systems.
Our results advance the exploration of integrability and manipulation of topological excitations in quantum critical systems.
\end{abstract}

\maketitle

\section{Introduction}

\begin{figure}
    \centering
    \includegraphics[width=0.75\linewidth]{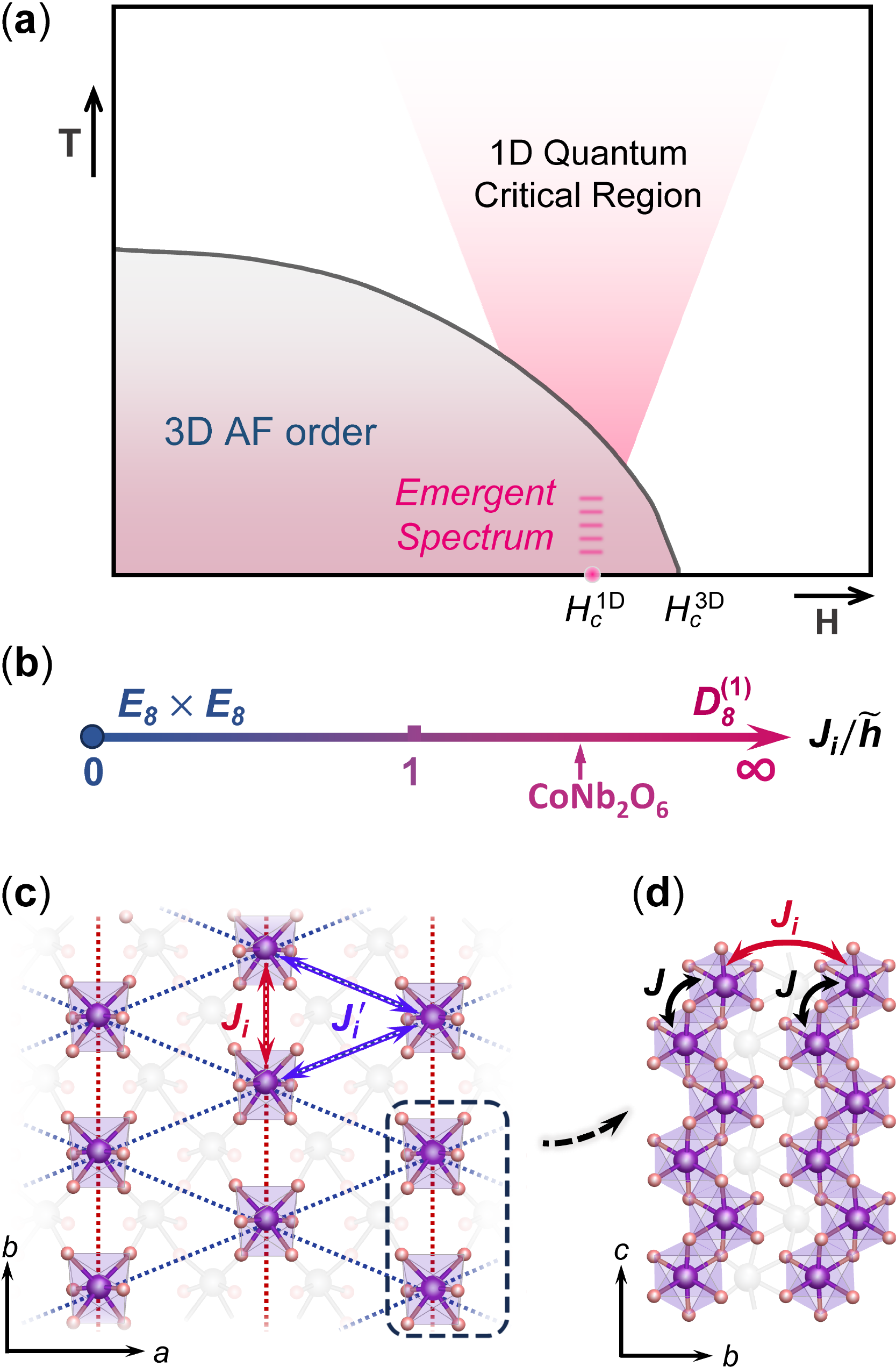}
    \caption{
    (a): Sketched phase diagram for \CNO~ under a transverse field ($H{\pll}b$), which consists of a 3D AFM phase at low temperatures and a prominent 1D quantum critical regime above the 3D ordering temperature. The (putative) 1D QCP at $H_c^{\rm{1D}}$ is hidden inside the 3D ordered phase close
    to the 3D QCP. The interchain interaction confines continuous critical excitations into gapped bound states that are characterised by the emergent \D81 Lie algebra.
    (b): Sketched phase diagram showing crossover between two integrabilities by tuning $J_i/\tilde{h}$ in the minimal model.
	(c) and (d): Illustration of the crystal structure of \CNO~ in the $a$-$b$ plane [in (c)] and
	in the $b$-$c$ plane [in (d)], showing Co$^{2+}$ ions (purple) inside the edge-sharing
	O$^{2-}$ octahedra (orange) and translucent coordinate Nb-O octahedra.
	The Co$^{2+}$ ions construct an isosceles triangular lattice with AFM interchain couplings $J_i$ and $J_i^{\prime}$ in the $a$-$b$ plane, and
	form a zigzag chain with FM intrachain coupling $J$ alone the $c$ axis.}
    \label{fig:cnopt}
\end{figure}

Enhanced quantum fluctuations near QCPs can give
rise to rich emergent phenomena~\cite{sachdev_2011,Coldea_2010,Senthil2004,Cui2023DQCP},
including enhanced symmetry, deconfined
fractional excitations, and emergent integrability.
As a prototypical model for studying quantum criticality,
the TFIC, which can be realised in a number of q1D quantum magnets, continues to be a research
hotspot~\cite{sachdev_2011,Coldea_2010,
Cui2019SCVO,WangZ2018,Sachdev2019,E8_2021,coldea2020,Morris2021,Liang2015,Xu2022,CoNbO2014PRX}.
In q1D magnets, the interchain couplings, though weak, are relevant perturbation that causes 3D magnetic ordering of the system. As a consequence, the genuine 1D QCP is usually hidden inside the 3D ordered phase as illustrated in Fig.~\ref{fig:cnopt}(a). This makes the critical behaviour even more complex and intriguing~\cite{CoNbO2014PRX,E8bcvoTHz}.
The significance of the 1D quantum
criticality in a TFIC is manifested by its spin excitation spectrum,
and in the 3D magnetic ordered phase,
the interchain interaction can be treated as
an effective weak longitudinal field $\tilde{h}$ that confines the
quasiparticles at critical into gapped bound states~\cite{E8_2021,xiao_2023}
with quantum \E8 integrability,
whose spectrum and scattering matrix are organised by the \E8 Lie algebra~\cite{ZAMOLODCHIKOV1989641,Coldea_2010}.

The above scenario for the emergent \E8 integrability was long predicted theoretically~\cite{ZAMOLODCHIKOV1989641} and the \E8 spectrum was initially proposed to be realised in the q1D Ising magnet \CNO~\cite{Coldea_2010}. Recently, the full \E8 spectrum and well defined dispersive \E8 quasiparticles have been observed in another q1D Ising magnet, BaCo$_2$V$_2$O$_8$, under transverse magnetic field~\cite{E8_2021,zhao_2021,xiao_2021,xiao_2023}. As for \CNO, however, results based on recent measurements are still controversial.

\CNO ~is a
renowned
q1D Ising magnet with zigzag ferromagnetic (FM) chains along the crystalline $c$ axis forming a frustrated isosceles triangular lattice in the $a$-$b$ plane as depicted in Fig.~\ref{fig:cnopt}(c) and (d). The FM intrachain coupling $J$ is
much stronger than the antiferromagnetic (AFM) interchain ones, $J_i$ and $J_i^\prime$
~\cite{CNOstruc1,CNOstruc2,CNOMAGstruc,interaction_refine}.
Experiments suggest that a 1D QCP in the TFIC universality class at $H_c^{\rm{1D}}\simeq 5.0-5.3$~T~\cite{Coldea_2010,Liang2015,Xu2022,CoNbO2014PRX,interaction_refine,zhe2020,Morris2021} is
hidden in the 3D
AFM ordered phase not far from the 3D QCP (at $H_c^{\rm{3D}}\simeq 5.5$~T~\cite{Coldea_2010}).
The seminal inelastic neutron scattering (INS) measurement~\cite{Coldea_2010}
provides a first evidence of \E8 spectrum:
The energy ratio of two lowest peaks identified is close to the golden ratio, which is the exact mass ratio of the two lightest \E8 particles,
when the system is tuned approaching $H_c^{\rm{1D}}$.
Recent THz spectroscopy measurements with much higher energy resolution,
however, revealed numerous additional satellite excitation modes that
surpass the \E8 description~\cite{zhe2020,Amelin_2022}. Besides, recent INS results indicate that the low-energy spectrum is influenced by the domain wall (DW)
interaction associated with the reduced lattice symmetry of \CNO~\cite{coldea2020}. These results
raise crucial questions about the adequacy of the Ising universality and the emergent \E8 spectrum.

In this article, we reexamine the spin excitations of \CNO~
near $H_c^{\rm{1D}}$ by performing iTEBD calculation and field theoretical analysis. We demonstrate that the low-energy physics in the vicinity of $H_c^{\rm{1D}}$ is controlled by the TFIC universality and the DW interaction is irrelevant. By suitably incorporating interchain spin fluctuations enhanced by spin frustration and the proximity to a 3D QCP, we find a remarkable result that the spectrum of \CNO~ is not consistent with the \E8 algebra but can be described by the \D81 Lie algebra associated with the Ising$_h^2$ integrable model. The satellite peaks incompatible with the \E8 algebra in the THz measurements are re-identified as the \D81 soliton and/or breather excitations. The topological single soliton excitation, usually forbidden in other systems, is found to be robust over a finite field range. We propose that the \D81 spectrum can be realised in a broad class of q1D quantum magnets. These results expand the realm of emergent phenomena in quantum 
magnetic systems.

\section{A minimal model for \CNO}
We consider the following Hamiltonian for \CNO,
\begin{equation}
    \mathcal{H}=\mathcal{H}_{\text{chain}}+\mathcal{H}_{\text{ic}},
    \label{Eq1}
\end{equation}
where $\mathcal{H}_{\text{chain}}$ and $\mathcal{H}_{\text{ic}}$ include the intra- and inter-chain interactions, respectively. 
The intrachain Hamiltonian proposed to precisely describe the magnetic properties of \CNO~ reads as~\cite{coldea2020}
\begin{equation}\label{Eq2}
\begin{aligned}
    \mathcal{H}_{\text{chain}}&=J\sum_{j}\left[-S^{z}_{j}S^{z}_{j+1}-\varepsilon(S^{x}_{j}S^{x}_{j+1}+S^{y}_{j}S^{y}_{j+1})\right.\\
    &\left.+\lambda_{af}S^{z}_{j}S^{z}_{j+2}+(-1)^{j}\lambda_{dw}(S^{z}_{j}S^{y}_{j+1}+S^{y}_{j}S^{z}_{j+1})\right],\\
    &-g\mu_{0}H\sum_{j}S^{y}_{j}
\end{aligned}
\end{equation}
where the first two terms form a 1D XXZ model in the FM Ising limit, the third term refers to an AFM interaction between the next-nearest neighbouring (n.n.n.) spins,
the fourth term is the so called DW interaction associated with the two DW bound state excitations, and the last term comes from the applied transverse magnetic field (along the crystalline $b$ axis).
The AFM and DW terms originate from the zigzag geometry of the chain.
Following Ref.~\onlinecite{coldea2020}, we take
$J=2.7607$~meV, $\varepsilon=0.239$, $\lambda_{af}=0.1507$, $\lambda_{dw}=0.1647$, and $g=3.100$, which were shown to accurately describe the low-energy excitations of \CNO. We note that more refined parameters for high-energy and strong-field excitations were recently proposed~\cite{interaction_refine,interaction_refine2}.

It is believed that the dominant interchain interaction is also of Ising-type, so we consider the following Hamiltonian
\begin{equation}
    \mathcal{H}_{\text{ic}}=\sum_{j,\langle \alpha,\beta\rangle} J_{i} S^{z}_{j,\alpha} S^{z}_{j,\beta},
    \label{Eq:Hinterchain}
\end{equation}
where chain labels $\alpha$, $\beta$ run over n.n. chains.
Given the weak coupling $J_i$, $\mathcal{H}_{\text{chain}}$ is usually treated at a chain mean-field level when the system is inside the 3D ordered phase,
$\mathcal{H}_{\text{ic}}\approx -\tilde{h} \sum_{j} S^{z}_{j,\alpha}$, where $\tilde{h}=-J_i \sum_{\beta} \langle S^{z}_{j,\beta} \rangle$ is an effective longitudinal field
acting on a single chain from its neighbours.

However, the frustrated interchain alignment causes cancellation of the effective fields from neighbouring chains and the interchain fluctuations, which are further enhanced by the proximity to the 3D QCP, must be treated in a more proper way. 
We then adopt a cluster mean-field approximation.
Taking into account the two-sublattice nature of the 3D AFM order,
we pick up a minimal unit consisting of two n.n. chains. The interchain couplings between these two chains are treated
exactly, while the couplings to other chains are
considered at the chain mean-field level as described above. The Hamiltonian of the minimal model then reads as
\begin{equation}
\mathcal{H}_{\text{min}}=\sum_{m=1,2} \mathcal{H}_{\text{chain}}^{(m)}+J_{i}\sum_{j=1}^N S^{z}_{j,1}S^{z}_{j,2}-\tilde{h}\sum_{j=1}^N\sum_{m=1,2} S^{z}_{j,m},
\label{Eq:ladderH}
\end{equation}
where $m$ is the chain index, and $\tilde{h}$ refers to the effective longitudinal field introduced by the chain mean-field approximation. Note that the minimal model recovers to two decoupled Ising chains when $J_i=0$, and an \E8 integrability is expected in this case. We will show in the following that the system crosses over to a novel quantum integrable class
exhibiting the \D81 mass spectrum in the $J_i/\tilde{h}\rightarrow\infty$ limit [Fig.~\ref{fig:cnopt}(b)].

\section{Emergent integrabilities of the minimal model}
By comparing the experimental and calculated spectra and computing the critical exponents in the minimal model, we first show that the non-Ising terms, including the DM interaction, are irrelevant to the TFIC universality of the 1D QCP [see Supplemental Materials (SM)~\cite{Supplemental}].
Then we analyse the emergent integrabilities of the minimal model in Eq.~\eqref{Eq:ladderH}.
Without $\tilde{h}$, each chain at $H_c^{\rm{1D}}$ is
described by a $(1+1)$D conformal field theory (CFT) with central charge $c=1/2$.
Either a small $\tilde{h}$ field or a small coupling $J_i$ can 
push the ladder system away from criticality into
a field-induced or intrinsic ordered phase, respectively.
The former one has the system emerged
the \E8 integrability emerges~\cite{zam},
while the later one drives the system
approaching the Ising$_h^2$ integrability.
In the $\tilde{h}=0$ limit, the low-energy physics is described by the following Ising$_h^2$ field theory
\begin{equation}
    \mathcal{A}_{\text{field}}= \sum_{\alpha=1,2} \mathcal{A}_{c=1/2}^{(\alpha)} + \lambda \int {\rm d}x~\sigma^{(1)}(x) \sigma^{(2)}(x),
    \label{Eq:HamCFTs}
\end{equation}
where $\mathcal{A}_{c=1/2}^{(\alpha)}$ denotes the $c=1/2$ CFT action for the critical chain $\alpha$, $\sigma^{(\alpha)}(x)$ and $\lambda$ refer to the correspondences of $S^z_{j,\alpha}$ and interchain coupling in the continuous limit,
respectively. In the
scaling limit [$a\;({\text {lattice spacing}}), \lambda\rightarrow0$
with finite $\lambda/a$],
the theory exhibits emergent Ising$^2_h$ integrability
characterised by the \D81 Lie algebra~\cite{LECLAIR1998523}. In the most general case where both $\tilde{h}$ and $J_i$ are present, the mass spectrum of the minimal model in Eq.~\eqref{Eq:ladderH}
crosses over from \E8$\times$\E8 to \D81,
as illustrated in Fig.~\ref{fig:cnopt}(d).

As for \CNO, the ordered moment at $H_c^{\rm{1D}}$ should be tiny since $H_c^{\rm{1D}}$ is
very close to $H_c^{\rm{3D}}$. This, together with the frustrated alignment of chains,
suppresses the effective field $\tilde{h}$. We therefore expect $J_i\gg\tilde{h}$ in \CNO, so that its excitation spectrum near $H_c^{\rm{1D}}$ is characterised by the \D81 algebra. Note that this is different from the case of BaCo$_2$V$_2$O$_8$, whose 1D QCP is located deep inside the 3D ordered phase~\cite{E8_2021}. There, the chain mean-field theory is a good approximation such that the excitation spectrum is well described by the \E8 algebra.

\begin{figure}
    \centering
    \includegraphics[width=0.85\linewidth]{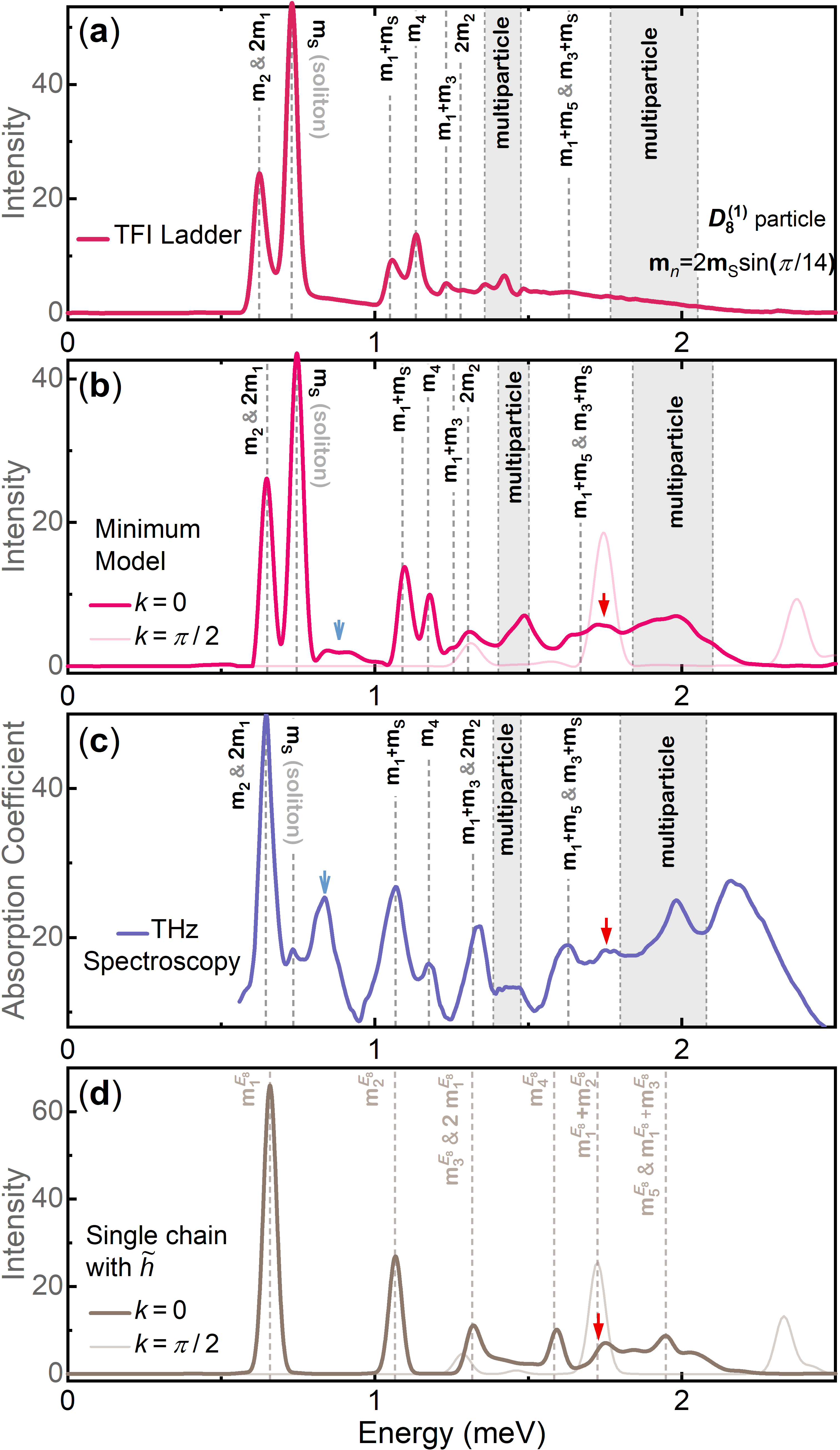}
    \caption{(a): Calculated zone-centre spectral functions at $H_c^{\rm{1D}}$ of an Ising ladder with $J=1$ meV, $J_{i}=0.36~J$ and $\tilde{h}=0$; (b): same as (a) but for the minimal model with $\lambda_{i}=0.1~J$ and $\tilde{h}=0$; (c):
    	THz absorption spectrum of \CNO~ at $H_{c}=5$~T and $T=0.25$~K, adapted from Ref.~\onlinecite{zhe2020}.
    (d): Calculated zone-centre spectral functions at $H_c^{\rm{1D}}$ of the single-chain model in Eq.~\eqref{Eq2} with the effective longitudinal
    field $\tilde{h}=0.034~J$, exhibiting the \E8 structure.
    In each panel, the vertical dashed lines at peak positions correspond to the masses of quasiparticles or bound states of multi-particles in the particular \D81 [in (a)-(c)] or \E8 [in (d)] model.
    Spectra at $k=\pi/2$ are also shown as light-coloured lines in panels (b) and (d) to demonstrate the zone-folding effect. Blue and red arrows refer to peaks associated with detailed microscopic Hamiltonian (see text). Note that many multiparticle modes are located in the shaded regimes, which give rise to multiple peak or plateau like spectrum.
    }
    \label{fig2}
\end{figure}

\section{Emergent \D81 spectrum of \CNO}
The integrable Ising$_h^2$ model in Eq.~\eqref{Eq:HamCFTs} possesses excitations associated with the \D81 algebra, which is characterised by a total of 8 types of particles, including one soliton ($S_{+1}$) and one antisoliton ($S_{-1}$), each with mass $m_{\text{s}}$, as well as 6 breathers $B_n$
with masses
$m_{n}=2m_{\text{s}}\sin(n\pi/14)$, $(n=1,\dots,6)$\;~\cite{LECLAIR1998523},
which are referred to \D81 particles in the following.
We expect that
peaks corresponding to these quasiparticles appear in the excitation spectrum of the minimal model at $\tilde{h}=0$ within
the energy-momentum range where the Ising universality dominates.
To check this, we numerically calculate the excitation spectrum at $k=0$ of
an Ising ladder with $J_i=0.36J$ at $H_c^{\rm{1D}}$. As shown in Fig.~\ref{fig2}(a),
we identify a series of peaks with energies precisely corresponding
to masses of \D81 particles. Besides the single particle peaks,
we also identify edges of several multiparticle excitation continua. Interestingly, the most prominent peak in the spectrum at about $0.74$ meV
originates from the unusual topological single (anti)soliton. Also note that
we cannot resolve any spectral signature corresponding to single $B_1$, $B_3$, or $B_5$ breather.
This verifies a recent theoretical prediction that these odd-parity breathers
cannot be excited from the ground state because of symmetry restriction~\cite{Gao2024}.
The agreement between the numerical and analytical results indicates that
the \D81 physics is robust even for a sizeable interchain coupling.

We then compare our result with the spectrum obtained in a recent
high-resolution THz measurement~\cite{zhe2020} for \CNO~ near $H_c^{\rm{1D}}$ (in Fig.~\ref{fig2}(c)).
Surprisingly, most peaks in the experiment can be assigned to single or multiple
\D81 particles, all the way up to about $2$ meV. To understand the two peaks (labelled by arrows) not captured by the Ising ladder model, we further calculate the spectrum of the minimal model
with the non-Ising interactions
and at $\tilde{h}=0$. As shown in Fig.~\ref{fig2}(b), peaks consistent with the
experiment emerge at arrow positions.
We find these
peaks are indeed associated with details of the microscopic model, instead of the \D81 algebraic structure. In fact, the peak labelled by the red arrow appears
at both $k=0$ and $k=\pi/2$, indicating it is a zone-folding peak caused by the n.n.n. AFM interaction $J_{\rm{AF}}$.
The peak at the blue arrow, however, is attributed to the confining effect from the DW interaction, which gives rise to additional bound states out of the $2m_1$ continuum of the \D81 spectrum (see Fig.~\ref{fig2}(a)).

\begin{figure}
    \centering
    \includegraphics[width=1\linewidth]{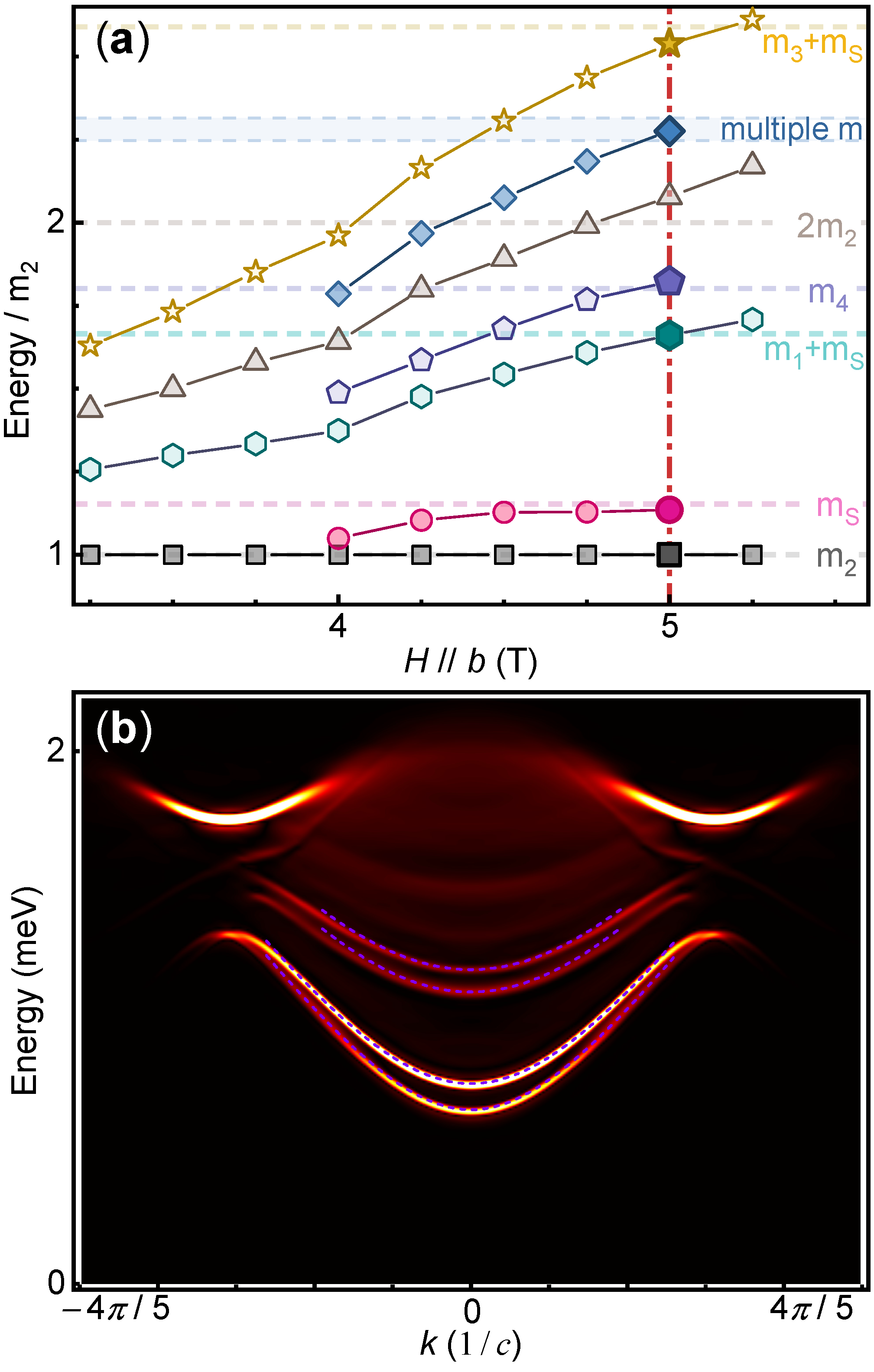}
    \caption{
    (a) Field dependence of several peak positions in the measured THz spectra extracted from Refs.~\onlinecite{zhe2020} and \onlinecite{Amelin_2022}. We take the lowest-energy peak (i.e. the $m_2$ of the \D81 at QCP)
    as the energy unit. The low-energy peaks fit to the \D81 algebraic structure best at about $H_{c}^{\rm{1D}} \simeq 5.0$~T.
    (b) The calculated spectrum of \CNO~ at $H_{c}^{\rm{1D}}$ in the entire BZ. The purple dotted lines are analytical predictions of several \D81 quasiparticle dispersions. }
    \label{fig3}
\end{figure}

We further calculate the spectrum of a single chain under an effective field $\tilde{h}$,
which is characterised by the \E8 algebra. As shown in Fig.~\ref{fig2}, many features
(satellite peaks) of the experimental spectrum are missing in the \E8 model but can be well captured by the \D81 one. In Fig.~\ref{fig3}(a), we plot the field dependence of several characteristic energies extracted from peaks
in experimental spectra presented in Refs.~\onlinecite{zhe2020} and \onlinecite{Amelin_2022}.
The energy ratios best fit to the \D81 mass spectrum at about $5$ T. (Putative) $H_{c}^{\rm{1D}}$ determined
in this way is very close to the value ($\sim 5.1-5.3$ T) determined from a previous NMR measurement~\cite{CoNbO2014PRX}. Note that by assuming the \E8 structure,
the same data gave $H_{c}^{\rm{1D}}\simeq 4.75$ T\;\cite{zhe2020} which deviated much to the NMR
value.
All these results unambiguously justify the emergent \D81 mass spectrum at low energies in \CNO.

As illustrated in Fig.~\ref{fig2}, Fig.~\ref{fig3}(a), and Ref.~\onlinecite{zhe2020},
with the transverse field increasing near $H_c^{\rm{1D}}$, one signature of the spectrum is
the splitting of \E8-like peaks where the \D81~ particles arise. Note that the mass ratio of $(m_1+m_s)/m_2\approx1.665$ in \D81 and that of $m_2/m_1\approx1.618$ in \E8~ are very close. Taking into account the field dependence of peak positions,
it is inadequate to assert any emergent integrability from calculating mass ratios of few low-energy peaks. An observation of a full spectrum is crucial.

From Fig.~\ref{fig3}(a), we find that the mass ratio $m_s/m_2$ varies
little over a finite field range near $H_c^{\rm{1D}}$, whereas the mass ratios of other \D81~particles show much stronger field dependence. This is because the single (anti)soliton, corresponding to a single domain wall, is a topological excitation that is robust against local perturbations. A single soliton
is usually forbidden.
Here it is inherent to the \D81~ algebra and stabilised as confined by the interchain coupling.

\section{Discussions and conclusions}
In Fig.~\ref{fig3}(b) we show the calculated spectrum of the minimal model
in the full BZ using the same parameters as in Fig.~\ref{fig2}(b).
 Clear dispersive bands corresponding to well defined \D81 quasiparticles
 show up in low energies. The bands
 span about 30\% of the BZ and follow
 the predicted relativistic dispersion of \D81 particle $\omega = \sqrt{m_{n}^2c^4+k^2c^2}$,
 where $m_n$ refers to the mass of the $n$-th quasiparticle, and $c$ is a characteristic velocity.
 Moreover, the masses follow scaling relation
 $m_n\sim J_i^{4/7}$ [Fig.~\ref{fig3}(c)] as predicted from the Ising$_h^2$ integrable theory (SM~\cite{Supplemental}).
 This is to be contrast to the \E8 model in which $m_n\sim J_i^{8/15}$.
 Interestingly, the dispersion and scaling relation apply to both the topological
 (anti)soliton excitation and the odd-parity breathers (which are invisible in the excitation spectrum)~\cite{LECLAIR1998523, Gao2024}.
 The unusual topological properties and symmetric restrictions
 make these particles potentially useful resources in quantum information.
 All these features deserve to be explored by spectral measurements such as INS and NMR.

Even though we do not perform fine tuning of model parameters,
our minimal model describes the spectrum
of \CNO~ quantitatively well. This suggests that our model, being
a cluster mean-field approximation, captures the correct emergent
low-energy physics of the system. We can understand this as follows:
Fixing the transverse field to
$H_c^{\rm{1D}}$, the interchain coupling is
definitely relevant when going from decoupled chains to a ladder. It
drives the system from critical to
inside the ordered phase
and causes emergent \D81 mass spectrum. We can further extend the cluster by including more chains, until it covers the full 3D system, where the approximation becomes exact. During this procedure, the system becomes non-integrable. But we expect that the interchain coupling only causes less relevant perturbation to the \D81 mass spectrum. This is indeed verified by our numerical calculation on 4 coupled Ising chains (Fig.~S5 of the SM~\cite{Supplemental}). Interestingly, we find that the spin frustration plays a crucial role in stabilising the \D81 spectrum. In the frustrated case, the spectrum resembles the \D81 of two decoupled ladders, whereas the spectrum turns to \E8 in the unfrustrated case.

Note that although we show the robustness of the \D81 mass spectrum for \CNO, the experimental
spectral intensity may crucially depend on both the microscopic details of interactions
and experimental setup, and is generically non-universal.
It is worth further noting that the emergent \D81 mass spectrum is not limited to \CNO,
but can be applied to a large class of quantum magnets. It obviously shows up in spectra of weakly coupled quantum Ising ladders with rung interaction much weaker than that along the ladder direction. Moreover,
we expect our argument for \CNO~ can be well applied to other
weakly coupled Ising chains when $H_c^{\rm{1D}}$ is close to $H_c^{\rm{3D}}$.
It would also be interesting to experimentally test the crossover from \E8 to \D81 physics by
tuning the distance of $H_c^{\rm{1D}}$ to $H_c^{\rm{3D}}$ via either rotating the field direction
or applying a pressure.
Last but not least, the physics should also be realised in specifically designed Rydberg atom systems.

In conclusion, we show that the spectrum of \CNO~ near its 1D QCP is described by the emergent \D81 algebra.
We show the emergent \D81 mass spectrum contains breather and exotic single-soliton excitations.
Our results advance the study on quantum integrability and topological excitations in quantum magnets.

\section{Acknowledgments}
We thank Linhao Li for helpful discussions.
This work is supported by the National Key R\&D Program of China (Grant No. 2023YFA1406500),
the National Natural Science Foundation of China (Grant Nos.~12334008, 12274288, and 12174441), the Innovation
Program for Quantum Science and Technology Grant No. 2021ZD0301900, and
Natural Science Foundation of Shanghai with Grant No. 20ZR1428400.

\bibliography{main}

\newpage
\appendix
\onecolumngrid

\section{Details of the iTEBD Simulations}
In this section we provide a concise overview of the process of calculating the dynamical structure factors (DSFs) in the infinite matrix product representation. Initially, we compute the zero-temperature space-time correlation  $\left\langle \hat{O}(0,0)\hat{O}(r,t)\right\rangle $ by employing the iTEBD method. The calculation of the space-time correlation of unitary operators is a routine operation within the iTEBD framework. This space-time correlation can be expressed as
\begin{equation}
\begin{aligned}
\left\langle \hat{O}(0,0)\hat{O}(r,t)\right\rangle &= \left\langle \psi_{G}\right|\hat{O}(0)e^{-it\hat{H}}\hat{O}(r)e^{it\hat{H}}\left|\psi_{G}\right\rangle \\
&=\left\langle \psi_{L}\right|e^{-it\hat{H}}\hat{O}(r)e^{it\hat{H}}\left|\psi_{G}\right\rangle.\label{eq:s1}
\end{aligned}
\end{equation}

In our study
the local observable is a local spin operator $S^\alpha\; (\alpha = x,\;y,\;z)$ which
is both Hermitian and unitary. The unitarity of the local spin operator guarantees the process $\left\langle \psi_{L}\right|=\left\langle \psi_{G}\right|\hat{O}(0)$ will not alter the canonical form of the ground state $\left\langle \psi_{G}\right|$.

Subsequently, standard operations of real-time evolution can be applied to the matrix product states $\left\langle \psi_{G}\right|\hat{O}(0)$ and $\left|\psi_{G}\right\rangle$. Then the dynamical structure factor of $\hat{O}$ can be determined by performing a Fourier transformation on $\left\langle \hat{O}(0,0)\hat{O}(r,t)\right\rangle $.

In practice, a sixth-order Suzuki-Trotter decomposition is utilised to minimise time-step errors. The values of the time step $\tau$ and the number of steps $N$ are 
optimised during the calculation. A larger $\tau$ results in a larger Trotter error and a narrower range of energy.  As $N$ increases, truncation errors also increase, and the memory cost grows quadratically. However, taking a large number of steps can improve the energy resolution. To balance computing resources and acceptable errors, we choose a truncation dimension $D=160$ for the computation of a ladder, and $D=80$ for the computation of a single chain. We also set $\tau=0.6J^{-1}$ and $N=2000$ as the maximum number of steps. Note that with the sixth-order decomposition, this parameter set already generates more precise results than taking $\tau=0.04J^{-1}$ in a second-order decomposition.

\section{Comparison of experimental and calculated spectra}
\begin{figure*}[h]
	\includegraphics[width=\textwidth]{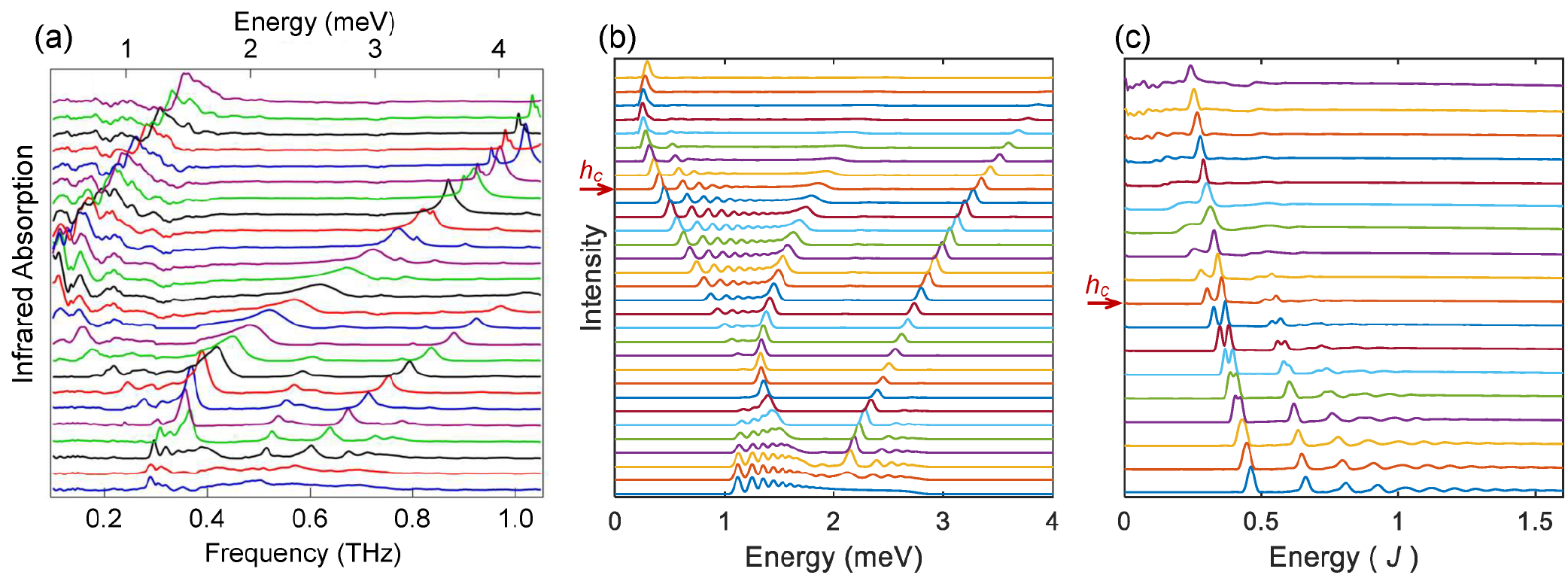}
	\caption{(a) The measured THz spectra of \CNO~ at 2.5~K with transverse field ranging from 0 to 12~T in 0.5~T steps, from bottom to top. Data adapted from Ref.~\onlinecite{Morris2021}.
		(b) The calculated spectra at $k=0$ of the single-chain model with a longitudinal field
		$\tilde{h}\simeq 0.01 J$ under transverse field ranging from 0 to 6~T in 0.2~T steps, from bottom to top.
		(c) The calculated spectra at the zone centre of the transverse field Ising ladder with $J_i=0.1J$ under transverse fields ranging from $0.42~J$ to $0.59~J$ in $0.01~J$ steps, from bottom to top.
	}
	\label{FigS3-spectra}
\end{figure*}

In this section, we present calculated zero-momentum spectra of the minimal model described by the Hamiltonian in Eq.~(4) of the main text in Fig.~\ref{FigS3-spectra}(b) and (c). For comparison, we also show the experimental spectra, adapted from Ref.~\onlinecite{Morris2021}, in Fig.~\ref{FigS3-spectra}(a).

We first show in Fig.~\ref{FigS3-spectra}(b) the spectra of a single chain under several transverse field values. Here the 3D ordering effect is treated by a longitudinal field $\tilde{h}$. This corresponds to setting $J_i=0$ in the minimal model of Eq.~(4). We can identify several characteristic peaks whose excitation energies increase with increasing the transverse field. Their
energy positions and field dependences are in accord with the measured ones, indicating that the single chain Hamiltonian in Eq.~(2) of the main text already captures main features of the spectrum of \CNO. However, we observe that several low-energy peaks in the experimental spectra split when increasing the field approaching the 1D QCP, $H_c^{\rm{1D}}$. This splitting is missed by the single chain model. To understand this splitting, we computed the low-energy spectra of an Ising ladder under several transverse fields near $H_c^{\rm{1D}}$, and the results are shown in Fig.~\ref{FigS3-spectra}(c). The spectra exhibit several different features to those of a single chain. First, the high-energy modes whose energy increase with increasing field disappear, implying that these modes are associated with the DW and other non-Ising terms of the microscopic Hamiltonian (see the next section for further discussions). Then, we focus on the low-energy excitation peaks. Away from $H_c^{\rm{1D}}$, they look similar to those of a single chain under an effective longitudinal field. With $H_c^{\rm{1D}}$ approaching, the low-energy peaks in the spectrum split and evolve with the field. Their energies and spectral weights redistribute and ultimately develop a structure following the \D81 algebra at $H_c^{\rm{1D}}$, which include soliton and breather excitations, as addressed in the main text. Upon further increasing the field, the single soliton peak persists at finite energies, while the peaks corresponding to breathers shift to lower energies and decay gradually. The behaviour of these low-energy peaks is highly similar to that observed in the experiment shown in Fig.~\ref{FigS3-spectra}(a). Similar behaviours of the spectra have been recently observed in several other
THz measurements in \CNO~\cite{Amelin_2022,Morris2021,zhe2020}.

It is worth noting that the critical field of the single chain in the calculation is slightly lower than the experimental value. This could be associated with either the absence of the interchain coupling in the calculation or calibration of magnetic field in experiments.
Nevertheless, this slight mismatch does not affect the overall field dependent behaviours of the spectra, especially for those gapped modes associated with the domain wall (DW) and other non-Ising interactions.

\section{Ising universality and Effects of the DW interaction}

\begin{figure}
	\centering
	\includegraphics[width=120mm	]{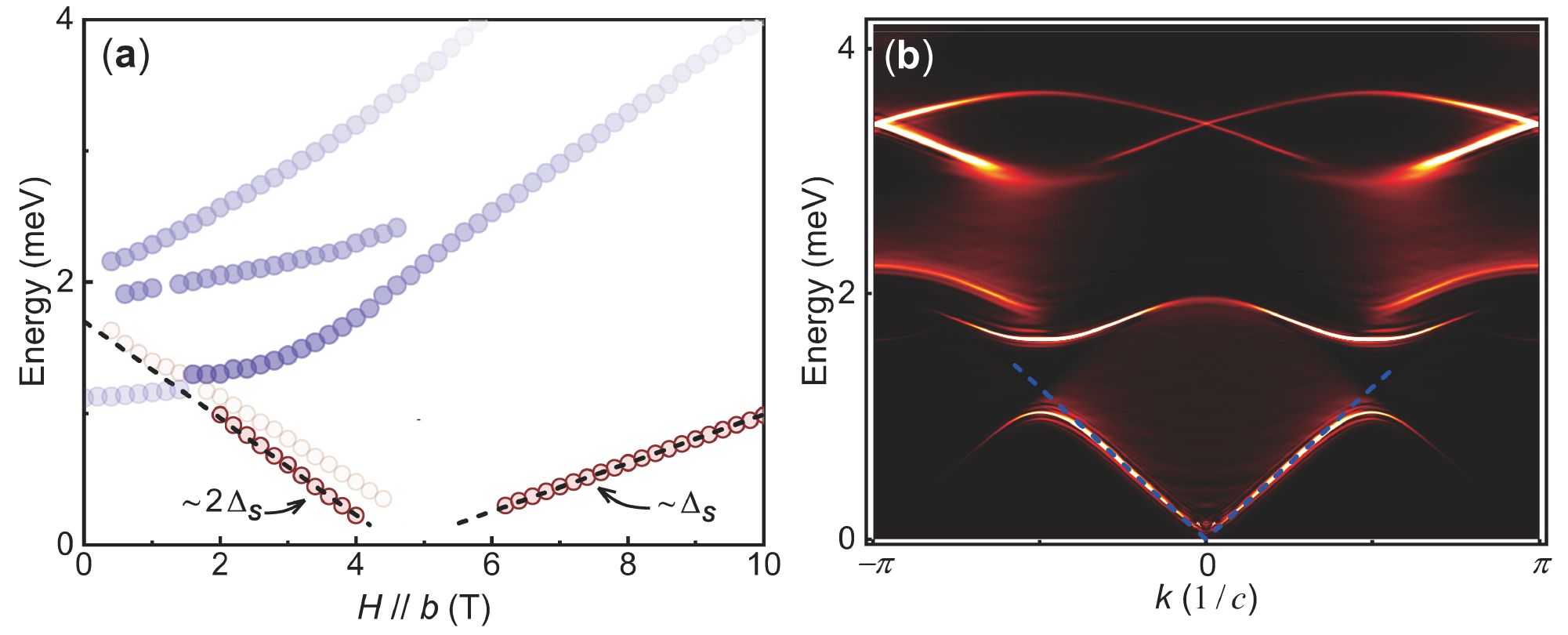}
	\caption{(a) Plot of the transverse field evolution of several characteristic energies extracted from peaks of spectral functions at $k=0$. Red points represent the lowest excitation mode of the Ising-type, which becomes critical at $H_c^{\rm{1D}}$. Deep-coloured points are calculated from the single chain model [Eq. (2) in the main text] without an effective longitudinal field $\tilde{h}$, and light-coloured ones are with $\tilde{h}$. Purple points represent modes associated the DW and other non-Ising interactions of the model. Dashed lines are guides to the eye with a factor of 2 difference in their slopes, in accord with the $(1+1)$D Ising universality. (b) The calculated dynamical structure factors of the single chain model at $H_c^{\rm{1D}}$ in the entire BZ. The Ising criticality is characterised by the linear dispersive mode guided by the blue dashed lines.}
	\label{figs22}
\end{figure}

As described in the main text, the microscopic Hamiltonian along the chain in Eq.~(2) contains several non-Ising terms. Here we show that these terms, especially the DW interaction, do not affect the Ising universality at $H_c^{\rm{1D}}$ (referring to the 1D QCP of a single chain, instead of the ladder). We have calculated the spectral functions at $k=0$ of a single chain under different transverse fields, described by Eq.~(2) [see Fig.~\ref{FigS3-spectra}(b)]. From peaks of the spectral functions, we can identify several characteristic energies. Their field evolution is shown in Fig.~\ref{figs22}(a). In a TFIC, the excitation gap decreases with increasing field in the ordered state. But the gap of the lowest energy mode initially increases with the field, as an effect of the DW interaction. Keep increasing the field, the lowest energy mode decreases after an avoid level crossing with another higher energy mode at about $H\simeq 2$ T. Approaching $H_c^{\rm{1D}}$,
the lowest excitation gaps for both $H>H_c^{\rm{1D}}$ and $H<H_c^{\rm{1D}}$ decrease linearly, and the gap ratio at the same distance to $H_c^{\rm{1D}}$ between the two sides is $2$.
These features imply the TFIC universality of the QCP is unaffected by the DW and other non-Ising interactions.

\begin{figure*}[h]
	\includegraphics[width=0.66\textwidth]{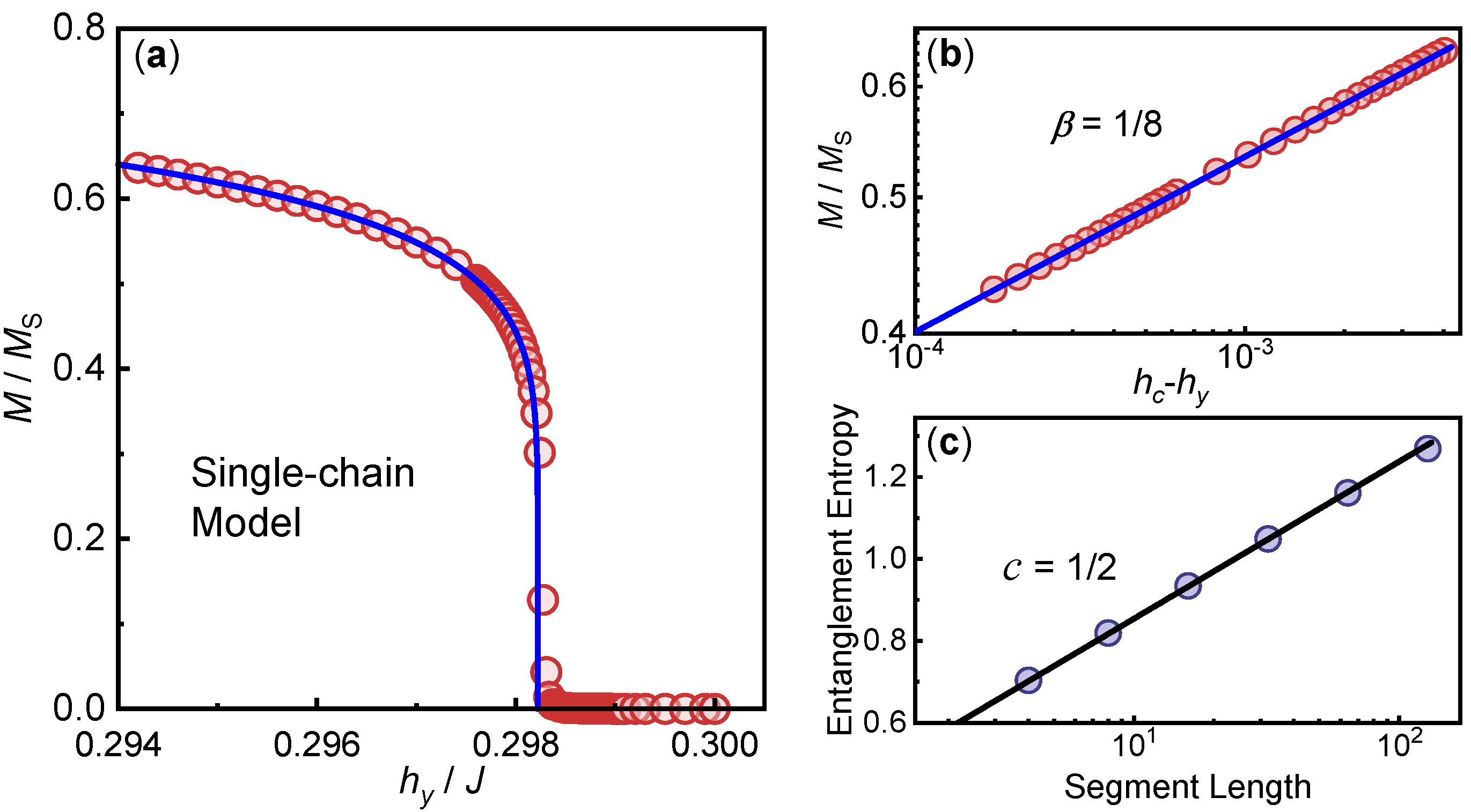}
	\caption{Results obtained via the iTEBD calculation on the single chain Hamiltonian in Eq.~(2) of the main text.
(a) The average of magnetisation $M$ versus the transverse field $H$. (b) Same data as in (a) but are plotted against $H_c^{\rm{1D}}-H$ on a log-log scale with the critical field determined to be $H_c^{\rm{1D}} = 0.29822$. The blue line is fit to the scaling function $M\propto(H_c^{\rm{1D}}-H)^{\beta}$ with the order parameter exponent $\beta=1/8$. (c) Entanglement entropy versus the length $L$ of the chain segment in the semi-log scale. The fitted slope value agrees with a central charge of $c= 1/2$. }
	\label{FigS1-single_scaling}
\end{figure*}

We then provide more evidences for the TFIC universality by examining the critical behaviours of magnetisation and entanglement entropy in the vicinity of $H_c^{\rm{1D}}$. As shown in Fig.~\ref{FigS1-single_scaling}, the order parameter, the magnetisation $M$, scales with the transverse field as $M\sim (H_c^{\rm{1D}}-H)^{1/8}$. Moreover, the entanglement entropy scales with the length of the chain segment as $E\sim \frac{c}{3}\ln L$ with the central charge $c=1/2$. These results further confirm that the single-chain model belongs to the (1+1)D Ising universality class near $H_c^{\rm{1D}}$.

In fact, by including the interactions in Eq.~(2) term by term and compare the corresponding spectra, we can identify the modes associated with each term. We find that the modes whose energies increase with increasing transverse field as shown in Fig.~\ref{FigS3-spectra}(b) and Fig.~\ref{figs22}(a) are associated with the DW and other non-Ising terms of the Hamiltonian.
For $H\sim H_c^{\rm{1D}}$, the DW related modes become dominant in the spectrum above about $2$ meV, as shown in Fig.~\ref{figs22}(a). To see this more clearly, we show the spectrum at the QCP in the entire Brillouin zone (BZ) in Fig.~\ref{figs22}(b). At low energies, there is a linear dispersive mode enveloping a continuum, aligning with the prediction of the Ising model. Around $2$ meV, the spectrum exhibits significant folding and flattening (compared to the typical bandwidth of the Ising model $\sim 2J\simeq 5.4$ meV) due to the DW interaction.
These features are consistent with recent THz and INS measurements
~\cite{Morris2021,Amelin_2022,interaction_refine,coldea2010}.

\section{Scaling of masses of the \D81~ and \E8~ quasiparticles}
\begin{figure*}[h]
	\includegraphics[width=0.5\textwidth]{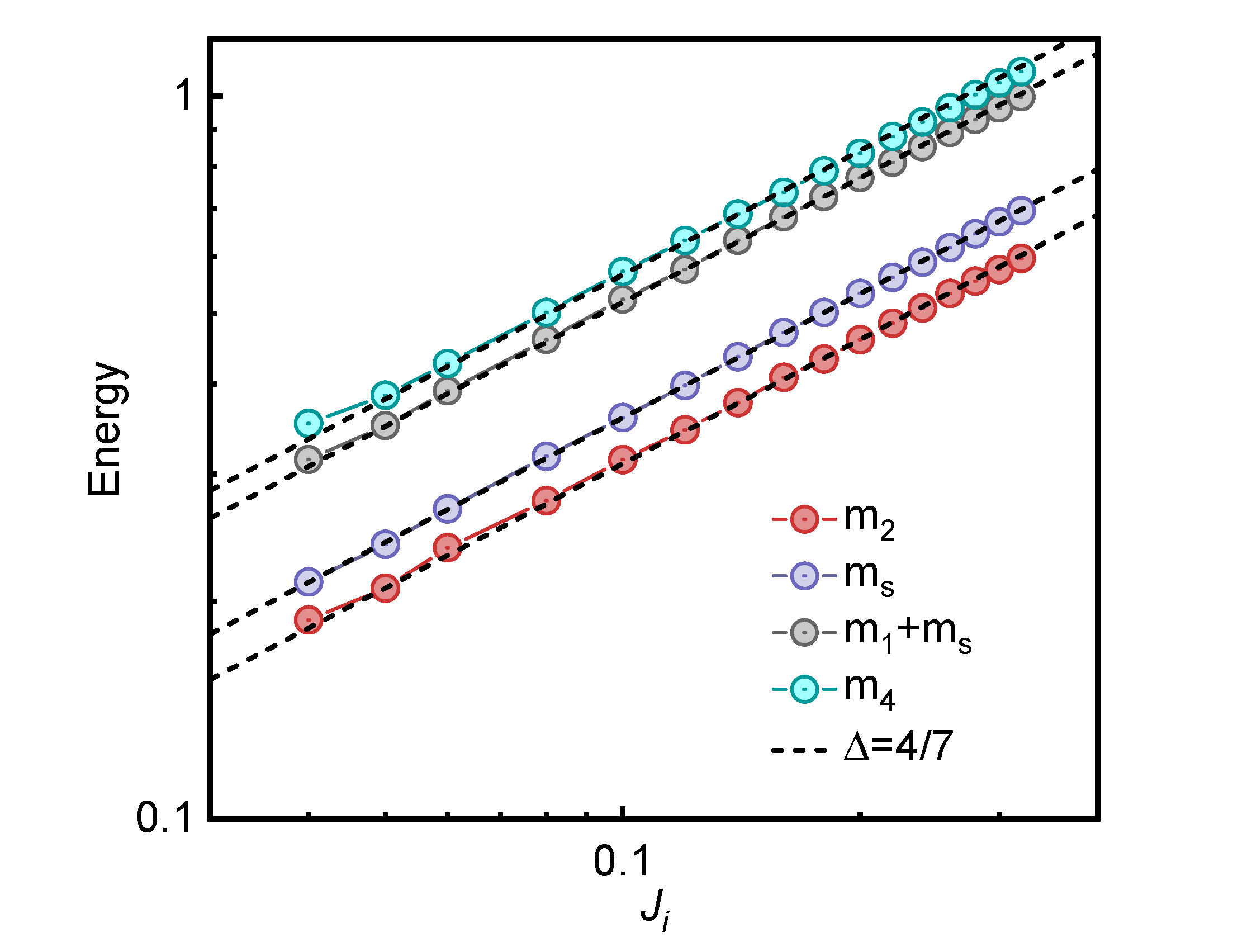}
	\caption{Scaling behaviours of several low-energy excitation modes in the spectrum at $k=0$ with the interchain coupling $J_i$ at $H_c^{\rm{1D}}$ in an Ising ladder ($\tilde{h}=0$) and a single chain ($\tilde{h}\propto J_i$), respectively. The extracted scaling exponents, $4/7$ in the ladder model and $8/15$ in the single chain, are consistent with the predicted values from the  Ising$_h^2$ and \E8~ integrable models.
}
	\label{FigS2-energy_scaling}
\end{figure*}
According to the Ising$^2_h$ quantum integrable field theory \cite{coupleCFT},
a standard scaling applies to masses of the \D81~ particles
\begin{equation}
m_{n}\sim \lambda ^{1/(2-d)},
\end{equation}
where $\lambda$ is the effective interchain coupling, and $d=1/4$ is the scaling dimension of the interchain interaction.
This leads to $m_{n}\sim \lambda^{4/7} \sim J_i^{4/7}$ for all \D81 particles including both soliton, anti-soliton, and breathers.

A similar scaling relation,
\begin{equation}
m_{n}\sim \tilde{h} ^{1/(2-d)},
\end{equation}
holds for \E8~ particles with $d=1/8$ \cite{DELFINO1995724}, where according to the chain mean-field approximation, the effective field $\tilde{h}\propto J_i$. Therefore, we expect that $m_{n}\sim J_i^{8/15}$.
In Fig.~\ref{FigS2-energy_scaling}, as well as Fig.~3(c) of the main text, we show the $J_i$ dependence of energies of several low-energy excitations at zone centre in the transverse field Ising ladder, which correspond to $m_2$, $m_s$, $m_1+m_s$, and $m_4$, respectively. All these modes show the same scaling behaviour consistent with the theoretical prediction. This scaling property provides further evidence in support of the \D81 mass spectrum, and can be experimentally detected. We also show in Fig.~\ref{FigS2-energy_scaling} the scaling behaviour of several \E8~ quasiparticles for comparison. As expected, they fall into a different scaling, with the power-law exponent $8/15$. This difference in the scaling behaviour can be used to determine the exact nature of the emergent integrability.

\section{Spectra of 4 weakly coupled Ising chains}
\begin{figure*}[h]
	\includegraphics[width=0.72\textwidth]{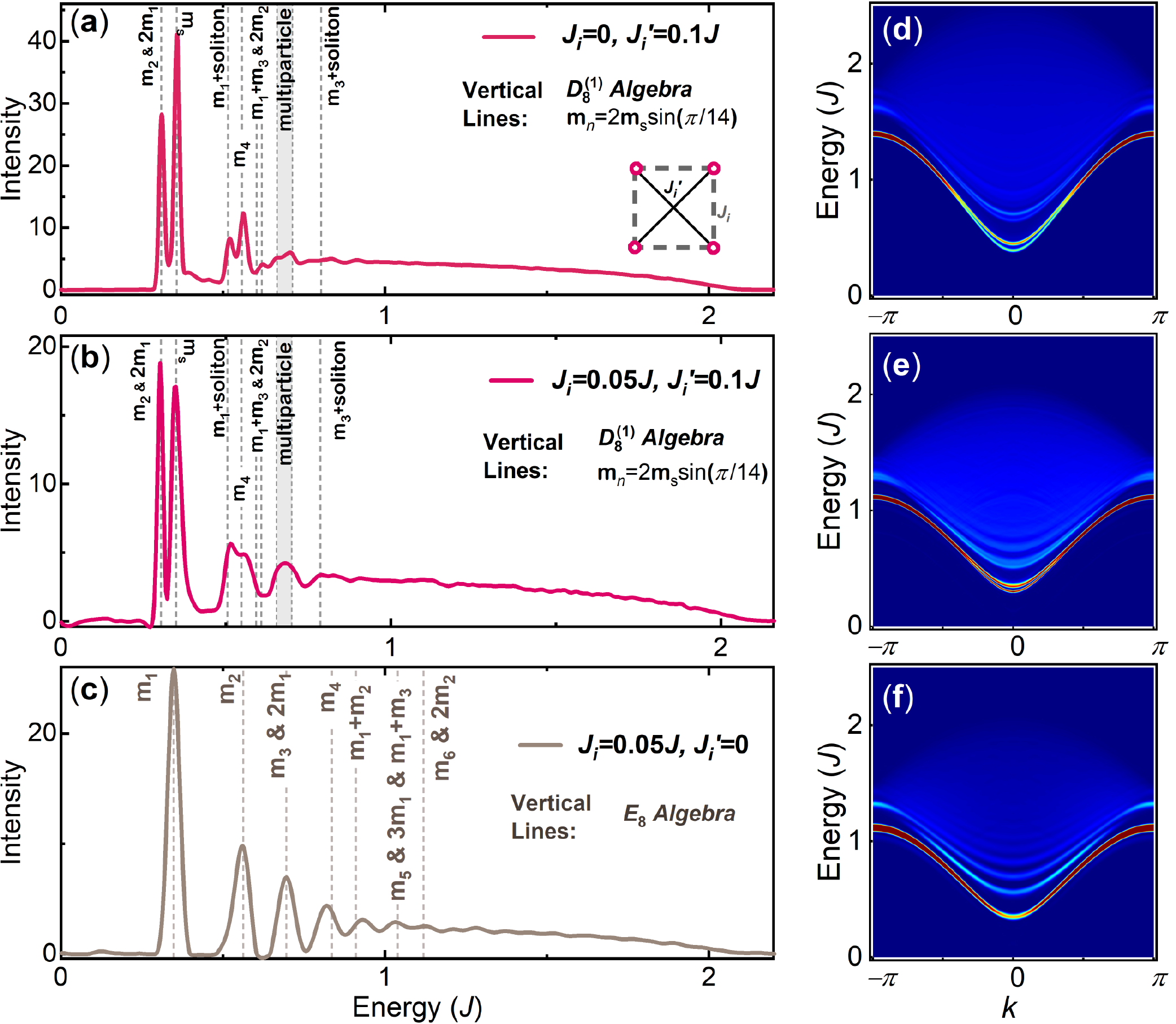}
	\caption{(a)-(c) Calculated spectra at $k=0$ and $H_c^{\rm{1D}}$ of 4 weakly coupled Ising chains [see inset of panel (a)] with
    (a) $J_{i}=0$ and $J_{i}^{\prime}=0.1~J$,
    (b) $J_{i}=0.05~J$ and $J_{i}^{\prime}=0.1~J$,
    and (c)  $J_{i}=0.1~J$ and $J_{i}^{\prime}=0$.
    In each panel, the vertical dashed lines at peak positions correspond to the masses of quasiparticles or edges of multiparticle bound states. The spectra in (a) and (b) are well described by the \D81 algebra, whereas the one in (c) is consistent with the \E8~ model.
    (d)-(f) are corresponding spectra in the entire BZ.
    The inset of panel (a) shows a sketch of how the 4 Ising chains are aligned.
     $J_i$ and $J_i^{\prime}$ refer to the nearest-
and next-nearest-neighbour interchain Ising couplings, respectively.
	}
	\label{FigS4-fourleg}
\end{figure*}
In the main text, we have shown, via a cluster mean-field approximation,
that the spectrum of \CNO~ at $H_c^{1D}$ is well described by the \D81~ algebra.
As for the validity of this approximation, we argue that we can extend the cluster
by coupling more chains. Although the system becomes non-integrable, the interchain fluctuations
only cause less relevant perturbation to the \D81~ spectrum. To see this is a valid argument, we hereby present calculated spectra of 4 coupled Ising chains in Fig.~\ref{FigS4-fourleg}. The spectrum in panel (a) is for two decoupled ladders. We see it is identical to the one of an Ising ladder, which is precisely described by the \D81~ algebra. In panel (b), we consider the case where the 4-chain system is fully frustrated with interchain couplings $J^\prime_i=2J_i$. In this case, one can easily check that the effective longitudinal field $\tilde{h}$ applied on each chain is exactly zero. This system is non-integrable. However, as shown in Fig.~\ref{FigS4-fourleg}(b), the peaks still resemble the \D81~ mass spectrum up to about $E\sim J$. In panel (c), we consider the unfrustrated case. Interestingly, the spectrum resembles that of the \E8 model instead of \D81.

In the comparison we observe that the frustration plays a crucial role for the \D81~ spectrum. The frustrated alignment of chains causes cancellation of effective longitudinal field $\tilde{h}$. Once $\tilde{h}$ is suppressed, the interchain fluctuations only add a perturbation to the \D81~spectrum, as expected. However, in the unfrustrated case, the longitudinal field $\tilde{h}$ dominates and drives the system to the \E8 integrability.
As for \CNO, the chains are indeed aligned in a frustrated way, and together with the proximity to the 3D QCP, the longitudinal field $\tilde{h}$ is substantially suppressed. In this case, as we illustrated in the 4-chain case, the interchain fluctuations only add a weak perturbation to the \D81 mass spectrum at low energy.

\end{document}